\def\aa{A\&A }					
\def\apj{ApJ }					

\def\ep{\rm e^+}
\def\em{\rm e^-}

\documentclass[a4paper,twocolumn]{esapub2005}
\usepackage{epsfig,amssymb,multirow}
\usepackage{times,graphicx}
\usepackage[sort,numbers]{natbib}

\title{Transport of positrons in the interstellar medium}

\author{W.~Gillard}
\author{P.~Jean}
\affil{C.E.S.R, CNRS/Universit\'e Paul Sabatier, Toulouse, France}
\author{A.~Marcowith}
\affil{L.P.T.A. CNRS-UM2 Universit\'e Montpellier II, France}
\author{K.~Ferri\`ere}
\affil{L.A.T.T, CNRS/OMP, Toulouse France}

\begin{document}
\maketitle

\begin{abstract}
This work investigates some aspects of the transport of low-energy $\ep$ ($E_{\rm k}\lesssim1\rm~MeV$) in the interstellar medium (ISM). We consider resonance interactions with magnetohydrodynamic (MHD) waves above the resonance threshold. Below the threshold, collisions take over and deflect $\ep$ in their motion parallel to magnetic-field lines. Using Monte-Carlo
simulations, we model the propagation and energy losses of $\ep$ in the different phases of the ISM until they annihilate. We suggest that $\ep$ produced in the disk by an old population
of stars (SNIa, LMXB), with initial kinetic energies below 1~MeV, and propagating in the spiral magnetic field of the disk, can probably not penetrate the Galactic bulge.
\end{abstract}

\section{Introduction}
The map of the 511~keV $\em\ep$ annihilation emission, measured by the spectrometer SPI on board ESA's INTEGRAL observatory, indicates that $\ep$ annihilate mainly in the Galactic bulge,
with a bulge-to-disk flux ratio $B/D\simeq2\pm1$ \citep{knod05}. Under the assumption that $\ep$ annihilate near their sources, this ratio provides severe constraints on the $\ep$ sources.
The measured value of $B/D$ does not correspond to any known distribution of astrophysical objects. The annihilation emission from the disk may be due to $^{26}\rm Al$ and $^{44}\rm Ti$ ejected by massive stars, but the origin of $\ep$ in the bulge is still a mystery. \citet{jean06} suggested that $\ep$ produced in the bulge with energies lower than a few MeV do not escape and, therefore, annihilate
in the bulge. \citet{prantzos06} argued that a ballistic transport of $1\rm~MeV$ $\ep$ from the disk into the bulge via a regular dipolar magnetic field could explain a high $B/D$ ratio.

The transport of $\ep$ is still poorly understood, but, as all charged particles, $\ep$ are sensitive to the Galactic electro-magnetic field, the flows of interstellar matter and the interactions with gas particles. In this paper, we report on preliminary results of $\ep$ transport in the Galaxy. Section~\ref{sec:MHD} describes the effects of {interactions with Alfv\'en waves}. Section~\ref{sec:coll} presents the model of collisional diffusion. The implications of our model and the other possible processes of diffusion are discussed in Section~\ref{sec:disc}. 

\section{Interactions with Alfv\'en waves}\label{sec:MHD}
In this section, we restrict the present analysis to the interactions between $\ep$ and Alfv\'en waves, usually assumed to be the main transport agent in the ISM. The effects of $\ep$ interactions with magnetosonic waves are postponed to a future work.

Regardless of the origin of Alfv\'en waves, the strong damping of Alfv\'en waves above the proton cyclotron frequency implies an energy threshold ($K_{_{\rm QL}}$) to the resonance condition. In the case of $\ep$, this condition can be satisfied only if \citep{jean06}
\begin{equation}
	\label{eq:QLth}
	\gamma\beta\geq12.85\times10^{-3}\frac{B_{\rm\mu G}}{\sqrt{n_{\rm cm^{-3}}}}~~,
\end{equation}
where $\gamma$ is the $\ep$ Lorentz factor, $\beta$ the $\ep$ velocity divided by the speed of light, $B$ the magnetic-field strength and $n$ the total particle density of the considered phase.
When $\ep$ are in resonance with Alfv\'en waves, we use quasilinear theory to derive the diffusion coefficient, which enables us to estimate the distance travelled by $\ep$ (see Equation~7 in \citep{jean06}).

Note, however, that when the amplitudes of magnetic fluctuations become comparable to the strength of the mean magnetic field, quasilinear theory gives a crude description of $\ep$ transport. This particular aspect will be addressed in a future paper.

\section{Interactions with gas particles}\label{sec:coll}
Below the quasilinear energy threshold, the resonance condition is not satisfied. Therefore, we assume that $\ep$ propagate in a collisional regime. We calculate the distance travelled by $\ep$ using Monte-Carlo simulations which take into account the physics of $\ep$ interactions with gas particles and their associated cross sections \citep{butler62,gould89,wallyn94,CH01,guessoum05}. The present analysis is restricted to a neutral hydrogen gas ($n=n_{\rm HI}$) and a fully ionized hydrogen gas ($n=n_{\em}+n_{\rm HII}$). The effects of the ionization fraction and metallicity will be presented elsewhere. Since the $n$-dependence of the quantities of interest is known analytically, we may arbitrarily adopt $n=1\rm~cm^{-3}$.

The medium is assumed to be pervaded by a homogeneous magnetic field $\vec{B_{0}}$, directed along the $z$ axis and having a strength of $5\rm~\mu G$ (in agreement with the typical value measured in the Galactic disk). Each $\ep$ travels along magnetic-field lines and describes a helical orbit of radius equal to its Larmor radius. When a $\ep$ interacts with a gas particle, its trajectory is deflected and the $\ep$ loses a fraction of its energy or it annihilates. Each simulation is performed using a total number of $\ep$ ranging between $5\times10^3$ and $2\times10^4$. The initial pitch angle is chosen randomly according to an isotropic distribution in the half-space $z\geq0$. In this regime, the distance travelled along magnetic-field lines by $\ep$ with a given initial energy does not depend on the magnetic-field strength. In the next sections, the results of our simulations are presented for $\ep$ with an initial kinetic energy of $3\rm~keV$.

\subsection{Transport in a neutral medium}
In a neutral medium, $\ep$ lose energy through ionization and excitation of atoms and through elastic scattering with atoms. They annihilate with bound $\em$ either directly or after forming a positronium in flight by charge exchange. 
To estimate the distance travelled by $\ep$ in a neutral medium, we distinguish three periods during the $\ep$ lifetime:
\begin{enumerate}
	\item The slowing-down period.\\	
	When the kinetic energy of $\ep$ lies between the {\it quasilinear threshold} and the charge exchange threshold (6.8~eV in H), $\ep$ lose energy mainly through ionization and/or excitation, and $\simeq 98~\%$ of them annihilate in flight, in quite good agreement with \citep{guessoum05}. In Figure~\ref{fig:NeutralSD}, we show the final distribution of $\ep$ parallel to magnetic-field lines. This distribution is obtained when the $\ep$ energy reaches the charge exchange threshold or when $\ep$ annihilate. The distribution provides an estimate of the average distance travelled by $\ep$ along magnetic-field lines. It can be fitted by a Gaussian, using $\chi^2$ minimization, with an average distance $\langle z\rangle\simeq(0.886\pm0.005)/n_{\rm cm^{-3}}\rm~pc$ and a spatial dispersion $\langle z^2\rangle^{1/2}\simeq(0.707\pm0.003)/n_{\rm cm^{-3}}\rm~pc$, obtained after an average slowing-down time $\langle t\rangle\simeq86/n_{\rm cm^{-3}}\rm~yr$.
\begin{figure}[!ht]
	\centering
	\includegraphics*[scale=0.9]{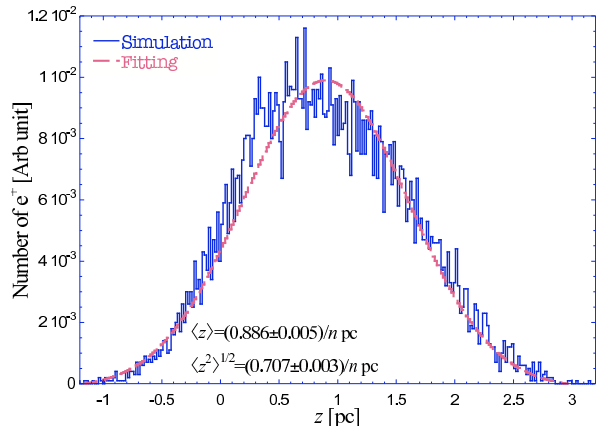}
	\caption{Spatial distribution of $\ep$ parallel to magnetic-field lines at the final stage  of the slowing-down period. The initial $\ep$ kinetic energy is $3\rm~keV$ and the total gas density is $n=1\rm~cm^{-3}$.}
	\label{fig:NeutralSD}
\end{figure}
	
	The distance travelled by $\ep$ during this period does not depend on the temperature of the gas, since their kinematic energy is always larger than the thermal energy of the gas ($\lesssim1\rm~eV$).
	
	Note that, in a neutral medium, Alfv\'en waves could be strongly damped by ambipolar diffusion. In this case, the slowing-down period of collisional diffusion would start at higher energies, maybe even at the injection energy of $\ep$. Starting collisional diffusion with an initial $\ep$ kinetic energy of $1\rm~MeV$ and using the theory of inelastic scattering \citep{GRYZ65}, we obtain  $\langle z\rangle\simeq4.0/n_{\rm cm^{-3}}\rm~pc$, $\langle z^2\rangle^{1/2}\simeq7.4/n_{\rm cm^{-3}}\rm~pc$ and an averaged slowing-down time $\simeq220/n_{\rm cm^{-3}}\rm~yr$.
	\item The thermalization period.\\		
	$\ep$ below the energy threshold for charge exchange lose energy elastically until their kinetic energy drops to that of gas particles. We obtain the spatial distribution parallel to magnetic-field lines that is shown in Figure~\ref{fig:NeutralTH}. The thermalization time is $\simeq10^{4}/n_{\rm cm^{-3}}\rm~yr$. Using $\chi^2$ minimization, we can fit the curve by a Gaussian with an average distance $\langle z\rangle\simeq(4\pm6)\times10^{-3}/n_{\rm cm^{-3}}\rm~pc$ and a spatial dispersion $\langle z^2\rangle^{1/2}\simeq(0.19\pm0.01)/n_{\rm cm^{-3}}\rm~pc$. During this period, $\ep$ annihilation is possible, but negligible ($\leq1\%$). The effect of the gas temperature is also found to be negligible. Between $8000\rm~K$ and $10\rm~K$, the difference in ${\langle z^2 \rangle}^{1/2}$ is less than 8~\%. 
\begin{figure}[!bth]
	\centering
	\includegraphics*[scale=0.9]{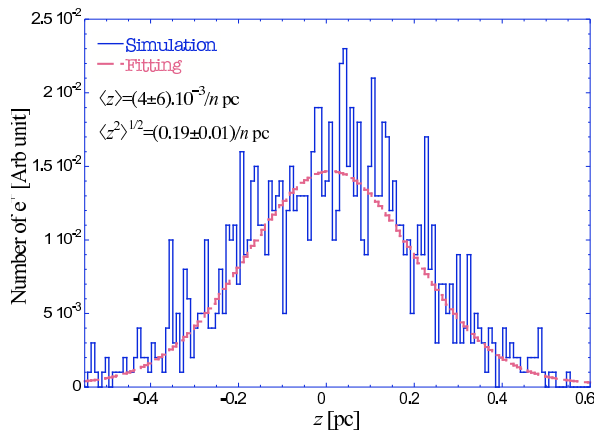}
	\caption{Spatial distribution of $\ep$ parallel to magnetic-field lines at the final stage  of the thermalization period. The initial $\ep$ kinetic energy is equal to the charge-exchange threshold and the total gas density is $n=1\rm~cm^{-3}$.}
	\label{fig:NeutralTH}
\end{figure}
	\item The thermalized period.\\	
	During this period, the kinetic energy of $\ep$ is comparable to the thermal energy of gas particles and $\ep$ scatter elastically with atoms. On average, there is neither gain nor loss in energy and $\ep$ diffuse over a distance $\lambda=\sqrt{2D\tau}$ along magnetic-field lines until they annihilate directly with bound $\em$ ($\tau$ is the $\ep$ lifetime \citep{guessoum05}). Given in Table~\ref{tab:DiffC} are the values of the diffusion coefficient $D$ parallel to magnetic-field lines that we computed for the different ISM phases. It has to be noted that both $D$ and $\tau$ depend on the gas temperature. 
\end{enumerate}
In agreement with the equation of motion of $\ep$ in a magnetic field, the magnetic-field strength influences only the perpendicular transport. In all three periods listed above, distances traveled perpendicular to magnetic-field lines are found to be negligible ($d_\perp\lesssim10^{-10}/n_{\rm cm^{-3}}\rm~pc$). 
\begin{table}
	\centering
	\begin{tabular}{lcc}
	\hline
	ISM phase & Temperature & Diffusion coefficient\\
	\hline
	Molecular     & 20~K & $2.1\times10^{21}/n\rm~cm^2.s^{-1}$\\
	Cold neutral & $100$~K & $1.2\times10^{21}/n\rm~cm^2.s^{-1}$\\
	Warm neutral & $ 8000$~K & $5.6\times10^{23}/n\rm~cm^2.s^{-1}$\\
	Warm ionized & $ 8000$~K & $4.4\times10^{17}/n\rm~cm^2.s^{-1}$\\
	Hot ionized & $10^6$~K & $7.8\times10^{22}/n\rm~cm^2.s^{-1}$\\
	\hline
	\end{tabular}
	\caption{Estimated values of the parallel diffusion coefficient in the thermalized period in the different ISM phases.}
	\label{tab:DiffC}
\end{table}

\subsection{Transport in an ionized medium} 
In an ionized medium, $\ep$ lose energy through Coulomb scattering with ions and free $\em$. They annihilate with free $\em$ either directly or after forming a positronium by radiative combination \citep{guessoum05}. Here, we model Coulomb interactions according to \citep{huba94} and \citep{butler62}. In contrast to propagation in a neutral medium, propagation here depends on the gas temperature. To estimate $\ep$ transport in an ionized medium, we  divide the $\ep$ lifetime into two periods:
\begin{enumerate}
	\item The slowing-down or thermalization period.\\	
	During slowing-down, $\ep$ lose energy through Coulomb scattering, and a negligible fraction ($\lesssim1\%$) of $\ep$ annihilate. In Figure~\ref{fig:IonSD}, we show the spatial distribution of $\ep$ parallel to magnetic-field lines at the end of the slowing-down period, i.e at a time $\sim25/n_{\rm cm^{-3}}\rm~yr$. The average distance $\langle z\rangle$ travelled by $\ep$ and the spatial dispersion $\langle z^2\rangle^{1/2}$ of the distribution are indicated in Table~\ref{tab:DistIon}, for both the warm ionized and the hot ionized phases.
	\begin{figure}[!th]
		\centering
		\includegraphics*[scale=0.9]{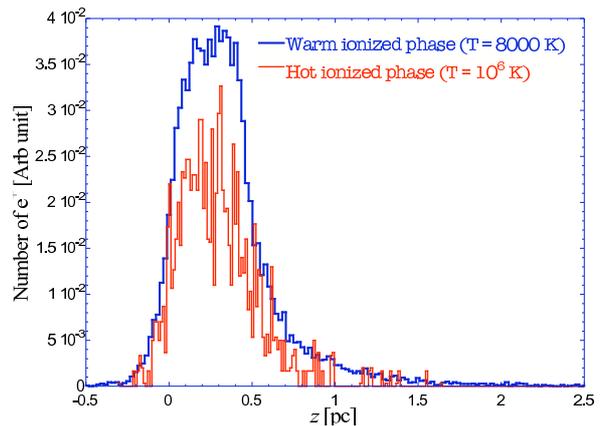}
		\caption{Spatial distribution of $\ep$ parallel to magnetic-field lines at the end of the slowing-down period, in the warm ionized (upper blue line) and the hot ionized (lower red line) phases. The initial $\ep$ kinetic energy is $3\rm~keV$ and the total gas density is $n=1\rm~cm^{-3}$}
		\label{fig:IonSD}
	\end{figure}
	\newpage
	\item The thermalized period.\\	
	During this period, the kinetic energy of $\ep$ is comparable to the thermal energy of gas particles and $\ep$ interact with $\em$ via Coulomb scattering. On average, there is neither gain nor loss in energy and $\ep$ diffuse over a distance $\lambda=\sqrt{2D\tau}$ along magnetic-field lines until they annihilate with free $\em$.  Here, too, the values of the diffusion coefficient $D$ parallel to magnetic-field lines are given in Table~\ref{tab:DiffC}.
\end{enumerate}
As in the case of a neutral medium, diffusion perpendicular to magnetic-field lines is found to be negligible ($d_{\perp}\lesssim2\times10^{-11}/n_{\rm cm^{-3}}\rm~pc$).\\
\begin{table}
	\centering
		\begin{tabular}{lcc}
		\hline
		ISM phase & $\langle z\rangle$ & $\langle z^2\rangle^{1/2}$ \\
		\hline
		Warm ionized & $0.34/n_{\rm cm^{-3}}\rm~pc$ & $0.35/n_{\rm cm^{-3}}\rm~pc$\\
		Hot ionized & $0.30/n_{\rm cm^{-3}}\rm~pc$ & $0.24/n_{\rm cm^{-3}}\rm~pc$\\
		\hline
	\end{tabular}
	\caption{Average distance $\langle z\rangle$ travelled by $\ep$ during the slowing-down period and spatial dispersion of the distribution $\langle z^2\rangle^{1/2}$ in the warm ionized and hot ionized phases. $n$ is the total gas density.}
	\label{tab:DistIon}
	\end{table}
\section{Discussion and conclusion}\label{sec:disc}

In this section, we discuss the effects of the transport of $\ep$ produced around $1\rm~MeV$ by an old stellar population (type Ia supernovae, low-mass X-ray binaries, ... ) in the Galactic disk. Table~\ref{tab:GalDiff} gives the average distances travelled by $\ep$ at different locations in the Galactic disk ($r\gtrsim3\rm~kpc$), as computed with the method presented in the previous sections. Because $\ep$ follow magnetic-field lines, we start by describing the magnetic-field properties in the disk.

\begin{table*}
	\centering
	\begin{tabular}{llcccc}
		\hline
		Galactic height & ISM phase & Total density $(\rm cm^{-3})$ & $K_{_{\rm QL}}$ (keV) & $d_{_{\rm QL}}\rm~(pc)$ & $d_{_{\rm Col}}\rm~(pc)$ \\
		\hline
		$z=0\rm~kpc$& Molecular & $(10^{2}~-~10^{6})$ & $\lesssim10^{-2}$ & $\lesssim 4$ & $\lesssim10^{-3}$ \\
		& Cold neutral & $(20~-~50)$ & $(5.3~-~2.1)10^{-2}$ & $(5~-~10)$ & $\lesssim10^{-2}$ \\
		& Warm neutral & $(0.2~-~0.5)$ & $(5.3~-~2.1)$ & $(50~-~80)$ & $(5~-~10)$ \\
		\hline
		$z=0.3\rm~kpc$ &Warm ionized & $\sim0.2$ & $\sim5.2$ & $\sim40$ & $\lesssim1$ \\
		& Hot ionized & $\sim0.002$& $\sim380$ & $\sim250$ & $\sim600$ \\
		\hline
	\end{tabular}
	\caption{Average distances travelled by $\ep$ with an initial kinetic energy of $1\rm~MeV$ in the quasi-linear $\left(d_{_{\rm QL}}\right)$ and collisional $\left(d_{_{\rm Col}}\right)$ regimes,  at different locations in the Galactic disk. The total particle densities are derived from \citep{ferriere98, ferriere01}}
	\label{tab:GalDiff}
\end{table*}
The current observational status, based on studies of the Galactic synchrotron emission \citep{beuer85} and of Faraday rotation measures \citep{simard80, inoue81, rand94, han99, han06}, is basically the following:\\
The interstellar magnetic field has uniform (large-scale) and turbulent (small-scale) components of comparable strengths ($\simeq 1.5~\mu$G and $\sim 5~\mu$G, respectively in the Solar vicinity).
The direction of the uniform magnetic field is nearly horizontal in most of the Galactic disk, and on the whole, the azimuthal component dominates. The uniform magnetic field probably has a spiral pattern, but it is not known whether this spiral is axisymmetric, bisymmetic, or a mixture of different azimuthal modes. The strength of the uniform magnetic field increases smoothly toward the Galactic center, reaching at least $4~\mu$G at $r \simeq  4$~kpc. Along the vertical, the uniform field strength decreases away from the midplane, probably following a two-layer structure with respective scale heights $\sim 200$~pc and $\sim 2~-~4$~kpc.


The old stellar population is more spread out along the vertical than neutral gas. Here, we assume for simplicity that at $z\sim0.3\rm~kpc$ the ISM is fully ionized. In accordance with the ISM distribution \citep{ferriere98, ferriere01}, we adopt $n\sim0.2\rm~cm^{-3}$ for the warm ionized phase and $n\sim0.002\rm~cm^{-3}$ for the hot ionized phase. With an initial kinetic energy $\sim 1\rm~MeV$, $\ep$ start propagating by diffusion on Alfv\'en waves until their kinetic energy falls below $\sim5.2\rm~keV$ in the warm ionized phase and $\sim380\rm~keV$ in the hot ionized phase. While $\ep$ are transported by Alfv\'en waves, they travel an average distance $\sim40\rm~pc$ in the warm ionized phase and $\sim250\rm~pc$ in the hot ionized phase. Below the resonance threshold, $\ep$ propagate by collisions until they annihilate. The average distance travelled in the collisional diffusion regime is $\lesssim1\rm~pc$ in the warm ionized phase and $\sim600\rm~pc$ in the hot ionized phase (see Table~\ref{tab:GalDiff}). Consequently, the maximum average distance travelled by $\ep$ before annihilation is $\sim850\rm~pc$. We also estimate that the maximum possible distance travelled by a typical $\ep$ is $\sim1\rm~kpc$. Assuming a spiral magnetic field in the Galactic disk, we then conclude that $\ep$ produced at Galactocentric radius $r\gtrsim 3\rm~kpc$ can not reach the Galactic bulge.
  
Note that in the current model, $\ep$ follow the average magnetic-field lines. However, in the Galactic disk, superbubbles and supernova remnants generate turbulence that can drive magnetic-field lines at high altitude ($\gtrsim1\rm~kpc$). There, $\ep$ are in a low-density medium where complex magnetic effects could significantly change the transport of $\ep$. Galactic winds could also transport $\ep$ advectively. In the Galactic center, a number of young massive stellar clusters blowing powerful winds have been observed \citep{ipavich75, portZ01}. Such points will be addressed in a future work. 

\section*{Acknowledgments}
A special thank to the local organizing committee of the $6^{th}$~INTEGRAL workshop.  W. Gillard would like to acknowledge J. P. Roques for the financial support.

\begin{thebibliography}{1}
{
	\bibitem[Beuermann et al.(1985)]{beuer85}
	Beuermann K., Kanbach G. and Berkhuijsen E. M., 1985, \aa~153:17
	
	\bibitem[Butler \& Buckingham(1962)]{butler62}
	Butler S. T. and Buckingham M. J., 1962, Phys Rev~126:1
	
	\bibitem[Charlton \& Humberston(2001)]{CH01}
	Charlton M. and Humberston J. W., 2001,  {\it Positron Physics}, in Cambridge Monographs on Atomic, Molecular and Chemical Physics
	
	\bibitem[Ferri\`ere(1998)]{ferriere98}
	Ferri\`ere K., 1998, \apj~497:759
	
	\bibitem[Ferri\`ere(2001)]{ferriere01}
	Ferri\`ere K., 2001, Rv Modern Phys.~73:1031
	
	\bibitem[Gryzinski (1965)]{GRYZ65}
	Gryzi\'nski  M., 1965, Phys~Rev~A~138:336
	
	\bibitem[Guessoum et~al.(2005)Guessoum, Jean \& Gillard]{guessoum05}
	Guessoum N., Jean P. and Gillard W., 2005, \aa~436:171
	
	\bibitem[Gould(1989)]{gould89}
	Gould R. J., 1989, \apj~344:232
	
	\bibitem[Han et al.(1999)]{han99}
	Han J. L., Manchester R. N. and Qiao G. J., 1999, MNRAS~306:371 
	
	\bibitem[Han(2006)]{han06}
	Han J. L., Manchester R. N., Lyne A. G., et al. 2006, \apj~642:868
	
	\bibitem[Huba(1994)]{huba94}
	Huba J. D., 1994, NRL/PU/6790D~94:265
	
	\bibitem[Inoue \& Tabara(1981)]{inoue81}
	Inoue M. and Tabara, H., 1981, PASJ~33:603
	
	\bibitem[Ipavich(1975)]{ipavich75}
	Ipavich F. M., 1975, \apj~196:107
	
	\bibitem[Jean et~al.(2006)]{jean06}
	Jean P., Kn\"odlseder J., Gillard W., et al. 2006, \aa~445:579
	
	\bibitem[Kn\"odlseder et~al.(2005)]{knod05}
	Kn\"odlseder J., Jean P., Lonjou V., et al. 2005, \aa~441:513
	
	
	\bibitem[Prantzos(2006)]{prantzos06}
	Prantzos N., 2006, \aa~449:878
	
	\bibitem[Portegies-Zwart et~al.(2001)]{portZ01}
	Portegies-Zwart S. F., Makino J., McMillan S. L. W., et al. 2001, \apj~546:L101
	
	\bibitem[Rand \& Lyne(1994)]{rand94}
	Rand R. J., Lyne A. G., 1994, MNRAS~268:497
	
	\bibitem[Simard-Normandin \& Kronberg(1980)]{simard80}
	Simard-Normandin M., Kronberg P. P., 1980, \apj~242:74
	
	
	\bibitem[Wallyn et~al.(1994)]{wallyn94}
	Wallyn P., Durouchoux Ph., Chapuis C., et al. 1994, \apj~422:610
}		
	
	
\end{thebibliography}

\end{document}